\documentclass[preprint2,10.5pt]{aastex}
\begin{document}

\title{\large{\rm{IDENTIFYING CONTAMINATED K-BAND GLOBULAR CLUSTER RR LYRAE PHOTOMETRY}}}
\author{D. Majaess$^1$, D. Turner$^2$, W. Gieren$^3$}
\affil{$^1$ Halifax, Nova Scotia, Canada}
\affil{$^2$ Saint Mary's University, Halifax, Nova Scotia, Canada}
\affil{$^3$ Universidad de Concepci\'on, Concepci\'on, Chile.}
\email{dmajaess@ns.sympatico.ca}

\begin{abstract}
Acquiring near-infrared {\it K}-band (2.2$\mu m$) photometry for RR Lyrae variables in globular clusters and nearby galaxies is advantageous since the resulting distances are less impacted by reddening and metallicity.  However, $K$-band photometry for RR Lyrae variables in M5, Reticulum, M92, $\omega$ Cen, and M15 display clustercentric trends. HST ACS data imply that multiple stars in close proximity to RR Lyrae variables located near the cluster core, where the stellar density increases markedly, are generally unresolved in ground-based images. RR Lyrae variables near the cluster cores appear to suffer from photometric contamination, thereby yielding underestimated cluster distances and biased ages. The impact is particularly pernicious since the contamination propagates a \textit{systematic} uncertainty into the distance scale, and hinders the quest for precision cosmology.  The clustercentric trends are probably unassociated with variations in chemical composition since an empirical {\it K}-band period-magnitude relation inferred from Araucaria/VLT data for RR Lyrae variables in the Sculptor dSph exhibits a negligible metallicity dependence: $(0.059\pm0.095)\times{\rm[Fe/H]}_{ZW}$, a finding that supports prior observational results. A future multi-epoch high-resolution near-infrared survey, analogous to the optical HST ACS Galactic Globular Cluster Survey, may be employed to establish $K$-band photometry for the contaminating stars discussed here.  
\end{abstract}

\section{{\rm \footnotesize INTRODUCTION}}
Photometry of RR Lyrae variables and stars in globular clusters is used to establish astronomical distances and the age of the Universe \citep{sz08,be11}. Reducing uncertainties associated with such photometry is a pertinent endeavour in the present era of precision cosmology, since even minute improvements to the distance scale can facilitate the breaking of degeneracies plaguing the selection of a cosmological model \citep{mr09}.   A mere $\sim0^{\rm m}.05$ systematic offset in distance for globular clusters arising from problematic photometry implies a potential $\sim5\%$ shift in derived ages for the clusters and the Universe.  Age estimates inferred for globular clusters set a crucial lower limit to the age of the Universe, while distances for the clusters and their RR Lyrae variable constituents are ultimately employed to constrain $H_0$, from which the age of the Universe stems.  Fundamental parameters established for RR Lyrae variables are used to anchor and calibrate other distance indicators \citep[e.g., the globular cluster luminosity function and Cepheid metallicity effect,][]{di06,ma10,fe11}, which emphasizes the importance of eliminating the propagation of systematic uncertainties into the cosmic distance ladder.

\begin{figure*}[!t]
\begin{center}
\epsscale{2.15}
\plotone{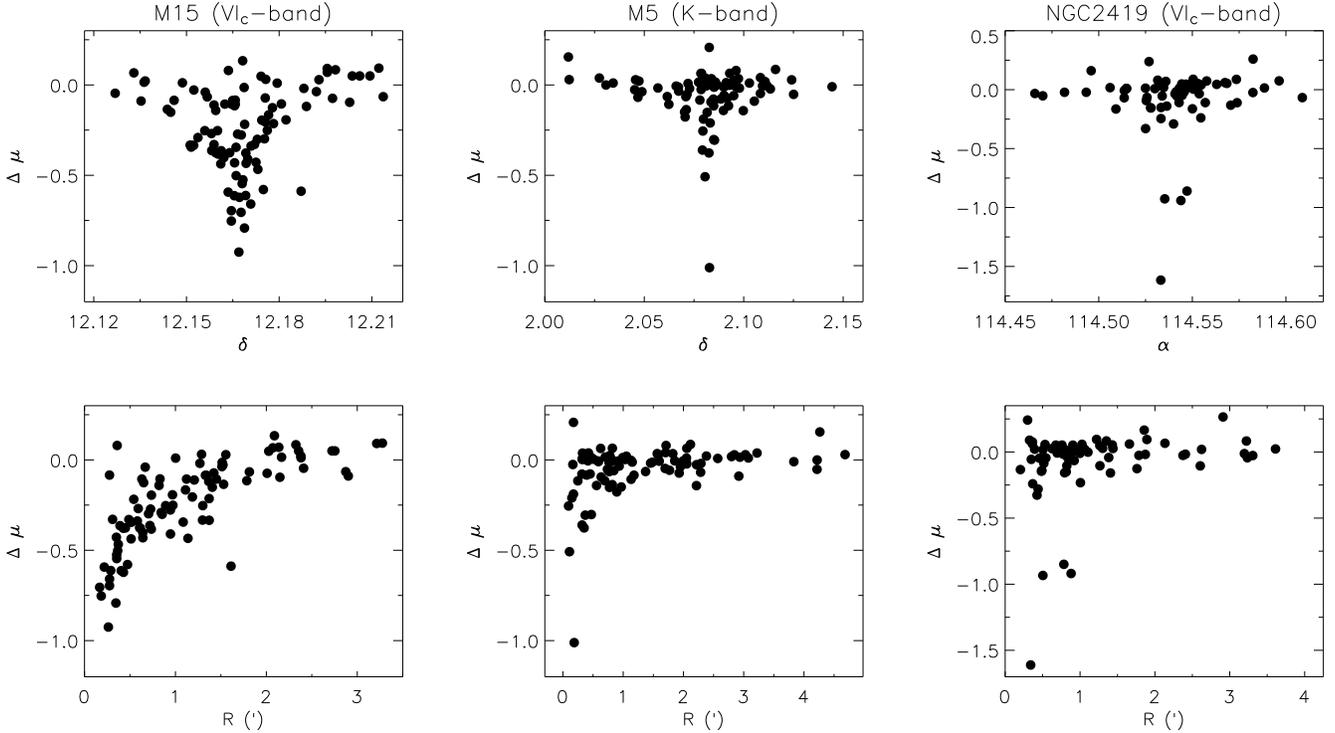}
\caption{\small{Wesenheit {\it VI}$_c$-band and {\it K}-band relative distance moduli computed for variables in M15, M5, and NGC2419. The moduli exhibit a dependence on position (i.e., declination, $\delta$) and clustercentric distance ($R$), whereby the distances appear underestimated for stars near the cluster core.  A similar dependence is noted for both $K$-band and reddening-free Wesenheit $VI_c$ distances.  That suggests that the former are not being severely impacted by differential reddening.}}
\label{fig-pos}
\end{center}
\end{figure*}

\begin{figure*}[!t]
\begin{center}
\epsscale{1.05}
\plotone{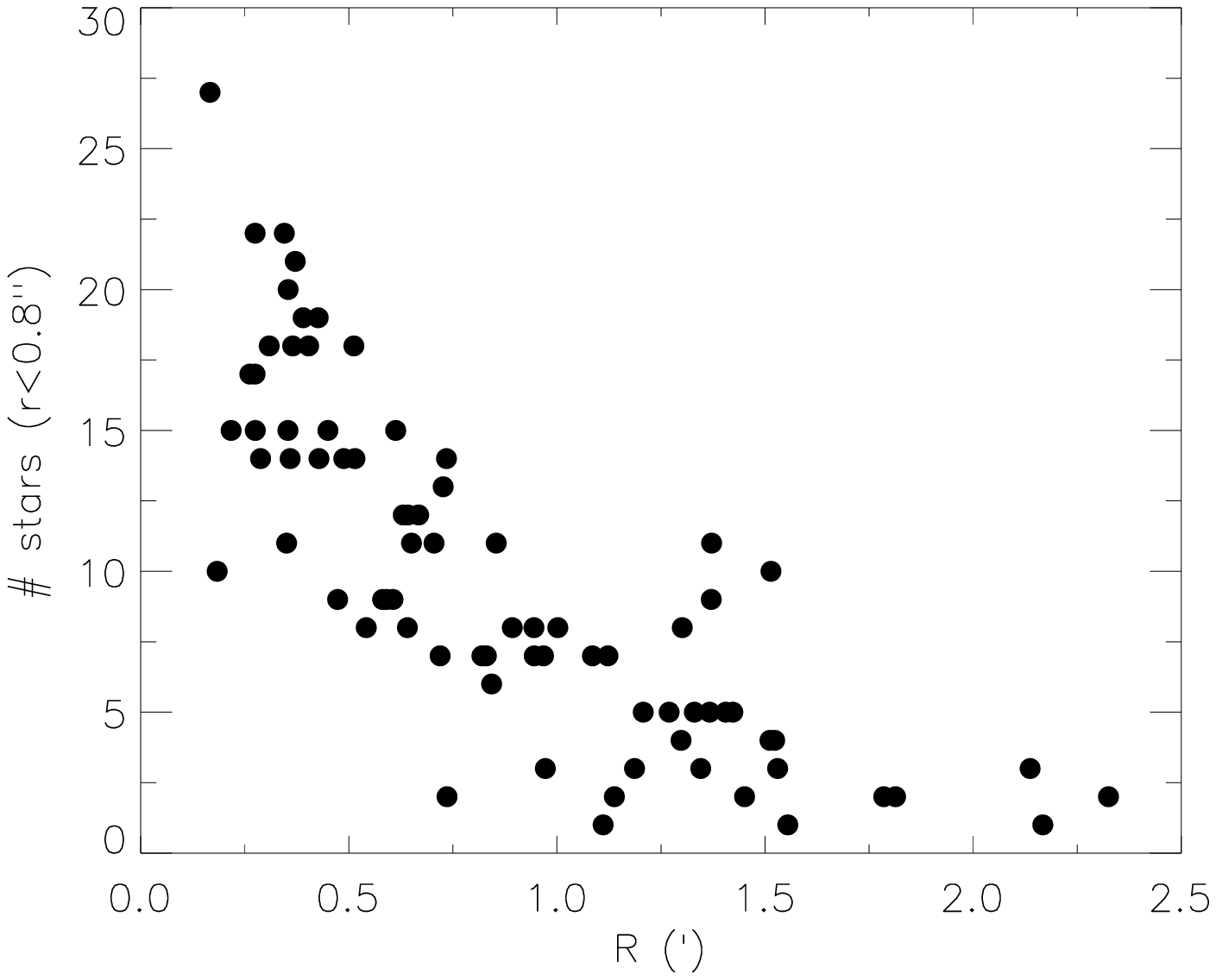}
\plotone{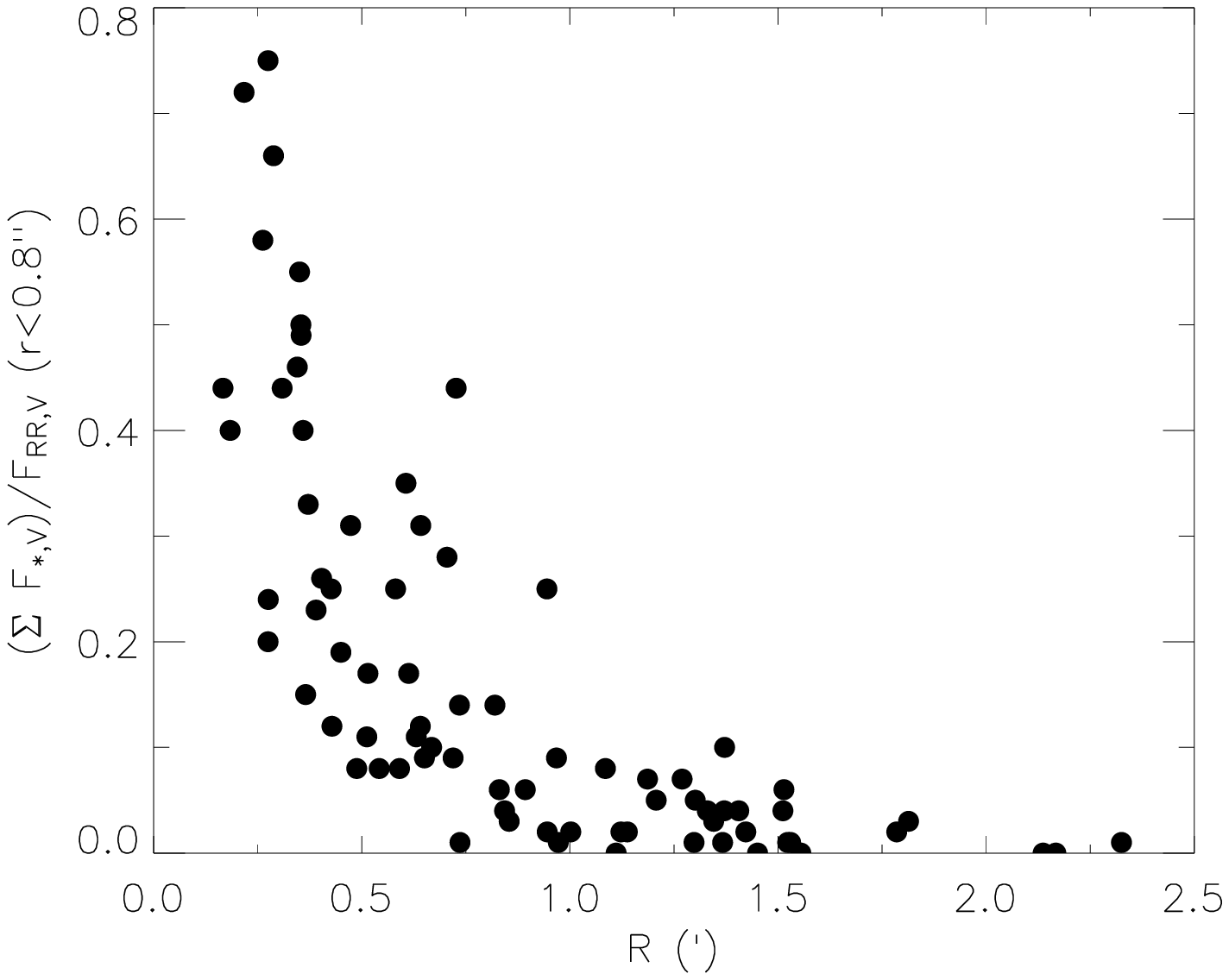}
\epsscale{0.7}
\plotone{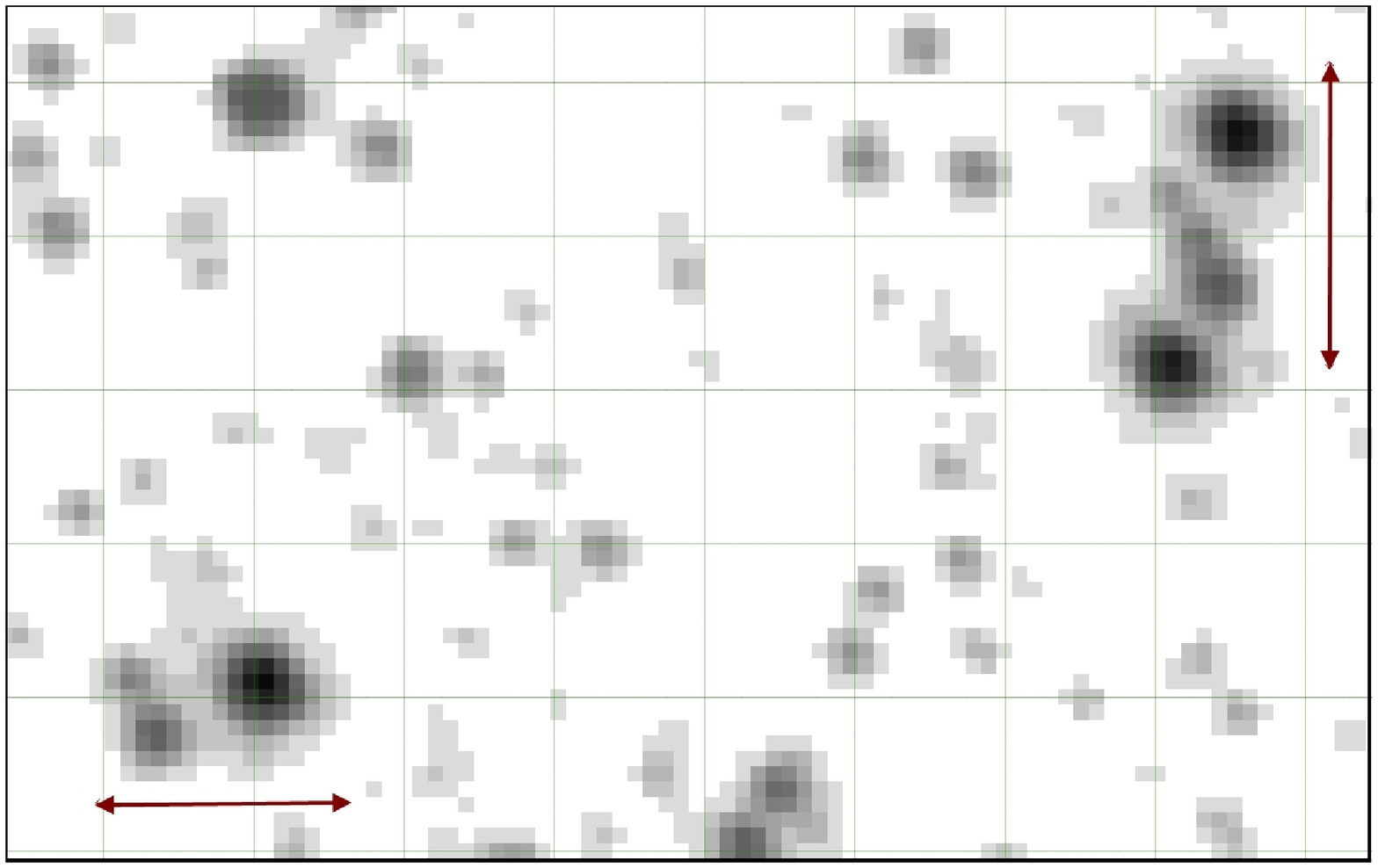}
\caption{\small{Top left, the density of HST ACS stars in M15 found $r<0.8 \arcsec$ (well within typical ground-based seeing) from the catalogued RR Lyrae variables increases with decreasing clustercentric distance.  Top right, the sum of the $V$-band flux for those neighboring stars divided by the mean RR Lyrae variable $V$-band flux ($\Sigma F_{*,V}/F_{RR,V}$) likewise increases with decreasing clustercentric distance.   Bottom, a cropped HST ACS image featuring two RR Lyrae variables near the core of M15. The RR Lyrae variables are the brightest stars in the two groups, and the red lines delineate $\sim1\arcsec$.  The HST ACS data imply that multiple stars lying in close proximity to RR Lyrae variables located near the cluster core are generally unresolved in ground-based images.  The flux measured for variables in the latter is subsequently contaminated by those stars.}}
\label{fig-acs}
\end{center}
\end{figure*}

 \citet{ma12} noted that Johnson-Cousins {\it VI}$_c$ photometry for RR Lyrae variables in several globular clusters (e.g., M15) exhibit a dependence on the (projected) clustercentric distance. The principal (yet not the sole) source giving rise to that trend is photometric contamination \citep[see also the discussion in][]{ya94,st05}. As the cluster core is approached the potential for photometric contamination increases in proportion to the stellar density, which rises rapidly. Indeed, \citet[][and references therein]{ma11b} asserted that blending/crowding compromises the Cepheid distance scale, since variables sampled near the cores of galaxies are typically brighter than their counterparts occupying the outer regions \citep[e.g., M101,][]{ke98}. However, the impact of blending on extragalactic variable star (i.e., Cepheids) distances is disputed, and many researchers instead favour metallicity differences as the origin of the brightening \citep[e.g.,][]{ss10,ge11}. Metallicity likewise increases with decreasing galactocentric distance \citep[e.g.,][]{ku12}, thereby introducing a degeneracy that needs to be overcome.

In this study the \citet{ma12} analysis is extended to the near-infrared ($\sim2.2\;\mu m$) regime, and the impact of photometric contamination is assessed for {\it K}-band photometry of RR Lyrae variables in the globular clusters M5 \citep{co11}, Reticulum \citep{da04,da05}, M92 \citep{dp05}, $\omega$ Cen \citep{so06}, and M15 \citep{lo90}.  The conclusions presented do not mitigate the importance of the studies discussed, and aspects of the RR Lyrae pulsation analysis remain unaffected.  For example, the derived pulsation periods place \textit{seminal} constraints on models since period change is a proxy for stellar evolution \citep{ku11,ju12}.   

\section{{\rm \footnotesize ANALYSIS}}
\begin{figure}[!t]
\begin{center}
\epsscale{.9}
\plotone{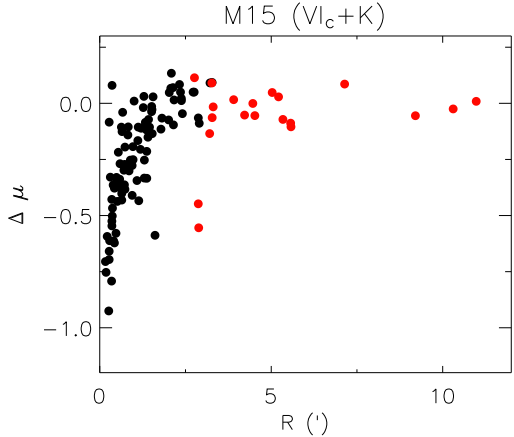}
\caption{\small{{\it K}-band (red dots) and Wesenheit $VI_c$ (black dots) relative distances for RR Lyrae variables in M15 plotted as a function of the clustercentric distance ($R$).  The $K$-band data were acquired through a large aperture \citep{lo90}, and thus contamination begins to appear at a larger clustercentric distance.}}
\label{fig-vik}
\end{center}
\end{figure}

In order to assess the impact of photometric contamination the residuals of the {\it K}-band period-magnitude relation are evaluated as a function of clustercentric distance. The canonical distance modulus for RR Lyrae variables is defined as:
\begin{eqnarray}
\nonumber
\mu_0 &=& K-M_K-A_K \\
\nonumber
M_K&=&\alpha \times \log{P_0}+\beta \\
\mu_0 &=& K-\alpha \times \log{P_0}-\beta -A_K
\end{eqnarray}
where {\it K} is the mean magnitude, $M_K$ is the corresponding absolute magnitude, $\log{P_0}$ is the fundamental mode period \citep[overtone pulsators are corrected using $\log{P_0}\simeq\log{P}+0.13$,][]{ca12}, $\beta$ is arbitrary since relative rather than absolute distances are evaluated to assess the contamination and stars in a given globular cluster are assumed to lie at the same distance, and $A_K$ accounts for dust extinction in the {\it K}-band. A period-Wesenheit ({\it VI}$_c$) analysis was adopted by \citet{ma12} to negate concerns that differential reddening might affect the clustercentric trends. The {\it VI}$_c$ Wesenheit function is comparatively insensitive to metallicity and also reddening free, but requires a photometric colour as input. The infrared photometry examined here is tied primarily to {\it K}-band observations, and the effects of differential extinction are not addressed directly save by demonstrating commonalities with the {\it VI}$_c$ Wesenheit trends (Figs.~\ref{fig-pos}, \ref{fig-vik}). Each RR Lyrae variable in a given cluster is therefore assumed to suffer the same (unknown) extinction $A_K$. If sizable differential reddening is present, it likely stems from patchy foreground dust distributed non-symmetrically across the cluster, and does not invalidate the broader conclusions drawn here. Moreover, the impact of differential reddening is minimized in the $K$-band since $A_K/E(B-V)\sim0.35$ \citep{ca89}, and indeed, there are numerous other advantages to employing $K$-band photometry \citep{bo03}.  Nonetheless, in general a reddening-free {\it VI}$_c${\it JHK} Wesenheit analysis is preferred and desirable to eliminate concerns regarding the impact of differential reddening for all clusters.   If $\beta$ and $A_K$ are chosen to be arbitrary, the above equation can be recast as:  
\begin{eqnarray}
\label{eqn-2}
\nonumber
\mu_0 &=& K-\alpha \times \log{P_0}-\beta -A_K \\
\mu_0 &=& K-\alpha \times \log{P_0}+\gamma 
\end{eqnarray}
The slope ($\alpha$) adopted for the period-$M_K$ function is approximately $-2.3$ \citep[][and references therein]{ca04,fe11}, and the conclusions derived below are not sensitive to that selection (to within the uncertainties). The residuals do not display a significant dependence on pulsation period.  

J2000 coordinates for the RR Lyrae variables are available via CDS and the compilations of \citet{cl01} and \citet{sa09}. For RR Lyrae variables discovered in Reticulum, J2000 coordinates were tabulated using the Aladin environment \citep{bon00} and the finder chart of \citet{wa92}.

In Fig.~\ref{fig-pos} relative $K$-band distances calculated for RR Lyrae variables in M5 are plotted as a function of position (i.e., declination, $\delta$) and clustercentric distance ($R$). Numerous stars in the cluster appear to suffer from contaminated photometry.  \citet{co11} remarked that the $K$-band photometry associated with numerous RR Lyrae variables in M5 was likely contaminated by neighboring stars \citep[see also][concerning blended objects in NGC2419]{di11}.   The extent, scale, and clustercentric dependence of that contamination is conveyed in Fig~\ref{fig-pos}.  The distance moduli are systematically underestimated for variables near the cluster core, where photometric contamination is most pronounced as a result of higher star densities (Fig.~\ref{fig-acs}).   PSF photometry does not correct for multiple stars coincident well within the FWHM of the system.  For example, higher-resolution HST photometry \citep{sar07} demonstrates that the density of stars lying within the seeing limit typically associated with ground-based images increases with decreasing clustercentric distance (Fig.~\ref{fig-acs}), and thus RR Lyrae variables located near the core are susceptible to increased contamination by neighboring stars.  The contaminating flux increases with decreasing clustercentric distance (Fig.~\ref{fig-acs}).   

Fig.~\ref{fig-pos} also features the relative distance moduli computed using a {\it VI}$_c$ Wesenheit function \citep{ma12} for RR Lyrae variables discovered in M15 and NGC2419. The results inferred from {\it K} and {\it VI}$_c$ photometry generally follow a similar trend, as demonstrated in Figs.~\ref{fig-pos} and \ref{fig-vik}, which in part suggests that differential reddening is not the cause of the trends noted in the $K$-band photometry.  However, it should be noted that {\it VI}$_c$-based relative distance moduli for RR Lyrae variables in M3 are anomalous \citep[][and references therein]{ma12}, and display an opposite trend to that inferred for the bulk of the clusters examined here and by \citet{ma12}. Variables near the core of M3 yield {\it VI}$_c$-based distance moduli that are overestimated relative to variables near the cluster periphery, for reasons that are not entirely clear. The acquisition of independent multi-epoch {\it VI}$_c${\it K} photometry for that sample is needed to help identify the source of the discrepancy.  

\begin{figure}[!t]
\begin{center}
\epsscale{1}
\plotone{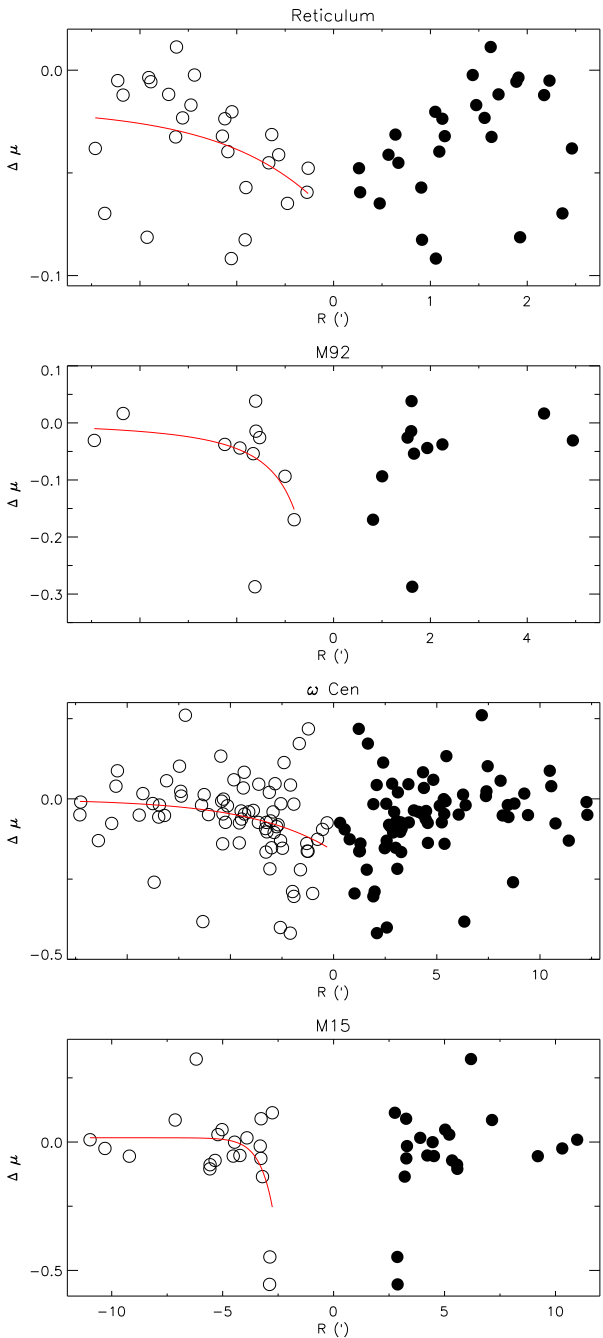}
\caption{\small{{\it K}-band relative distance moduli for RR Lyrae variables in Reticulum, M92,  $\omega$ Cen, and M15 are plotted as a function of the clustercentric distance ($R$). The data are mirrored (open circles) to better convey the clustercentric trends.  The best fit exponential functions are overlayed.}}
\label{fig-r}
\end{center}
\end{figure}

Relative distance moduli calculated for RR Lyrae variables in Reticulum, M92, and $\omega$ Cen are plotted as a function of clustercentric distance in Fig.~\ref{fig-r}. All clusters display radial trends whereby variables closest to the cluster core are generally brighter than objects near the periphery. The distribution of RR Lyrae variables in M5, $\omega$ Cen, M92, and M15 ($K$-band) is such that a less-biased distance may potentially be estimated by relying on stars near the cluster periphery (Figs.~\ref{fig-pos}, \ref{fig-r}).   However, the trends displayed for Reticulum present problems for deriving an unbiased distance. RR Lyrae variables in that cluster exhibit small random scatter, yet the extent of the pernicious systematic uncertainty is unclear (see also M15 $W_{VI_c}$ data, Fig.~\ref{fig-pos}). Supplemental observations are desirable.  Nevertheless, the distance offset between RR Lyrae variables located near the core and periphery may be estimated for M5, M15 ($K$-band), $\omega$ Cen, and M92.  The resulting estimated offsets are $\Delta \mu \sim-0.09,-0.13,-0.07,-0.06$ for the aforementioned clusters, respectively.  The differences are evaluated for illustrative purposes, and are subjective and uncertain owing partly to an inability to overcome poor statistics to assess the extent of the compromised photometry (e.g., M92 and M15 ($W_{VI_c}$ data)).  The bias noted may be reduced by remaining cognizant of the trends highlighted in Figs.~\ref{fig-pos} and ~\ref{fig-r} and applying robust estimators, among which median and sigma-clip algorithms are perhaps the simplest, rather than least-squares or average which are sensitive to outliers.  Indeed, the trends highlighted in Figs.~\ref{fig-pos} and ~\ref{fig-r} provide the rationale needed when eliminating certain outliers. 

Pearson coefficients computed for M5, M15 ($K$-band), Reticulum, M92, and $\omega$ Cen are: r = 0.35 (p = 0), r = 0.28 (p = 0.21), r = 0.36 (p = 0.07), r = 0.37 (p = 0.26), and r = 0.25 (p = 0.03), respectively.  Pearson coefficients range from $r=-1$ to $+1$ and indicate the strength of negative and positive correlations, respectively, while $r=0$ indicates a lack of correlation and $p$ represents the significance. The results are to be considered in tandem with the evidence hitherto tabulated, which in sum indicates that the $K$-band residuals for the clusters examined are correlated with the clustercentric distance.  However, caution is warranted when interpreting the coefficients.  Blending is radially dependent and non-linear, and the blending trends differ between clusters owing to the different radial sampling, seeing, core and apparent density, cluster distance, etc. (Fig.~\ref{fig-vik}).  The coefficients cited are an average for the entire cluster, which is typically weighted between the marginal to null correlation existing for stars near the cluster periphery, and the stronger correlation tied to stars near the core (Fig.~\ref{fig-pos}).

\begin{figure}[!t]
\begin{center}
\epsscale{1}
\plotone{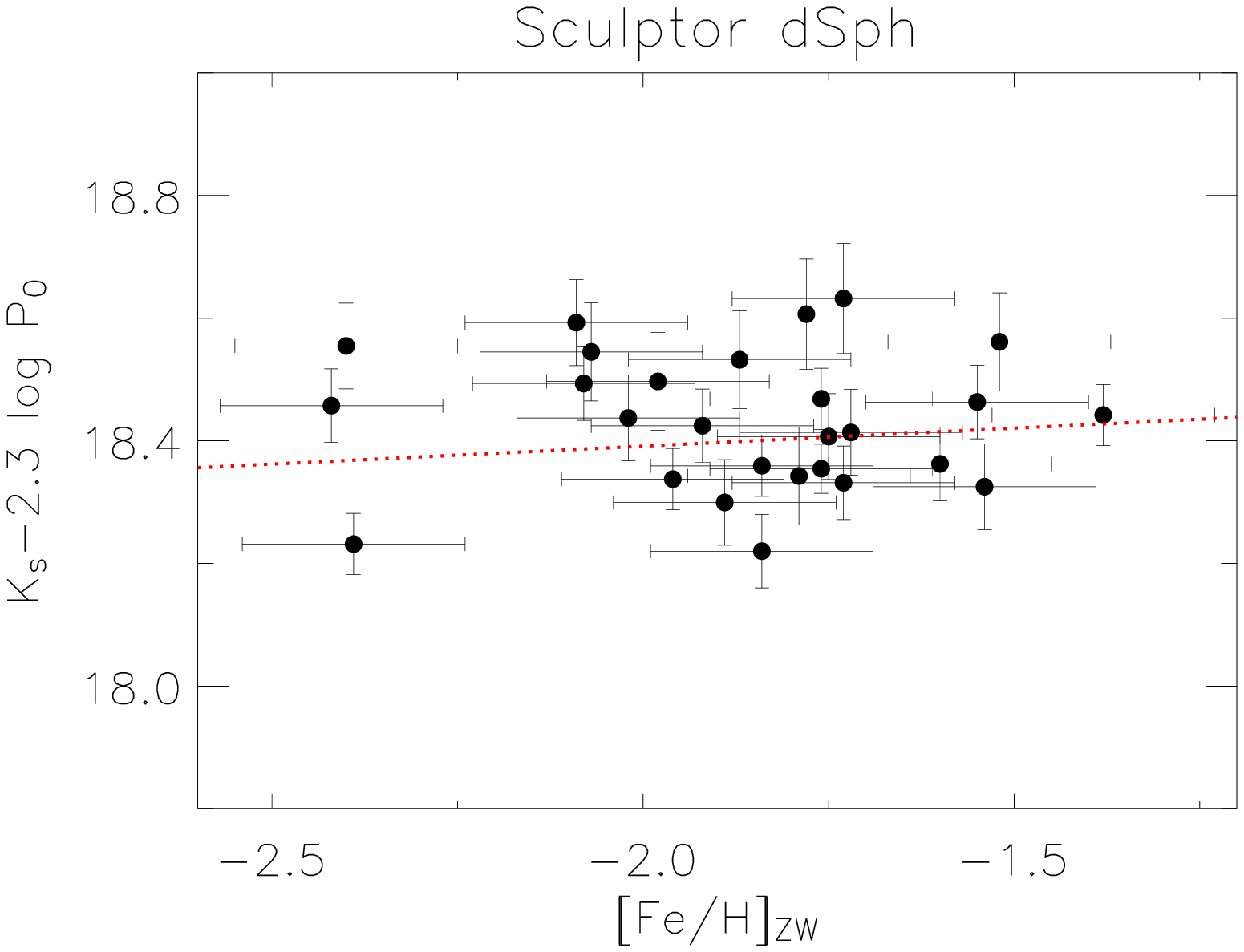}
\caption{\small{$K_s$ data for RR Lyrae variables in Sculptor \citep{cl05,pi08} are plotted as a function of the iron abudance. The best fit yields an insignificant slope of $(0.059\pm0.095)\times{\rm[Fe/H]}_{ZW}$, and implies that the $K_s$ observations are comparatively insensitive to metallicity.  The uncertainties may be mitigated by obtaining $K_s$ data for additional RR Lyrae variables in the Sculptor dSph that possess metallicity estimates \citep{cl05}.}}
\label{fig-met}
\end{center}
\end{figure}

\subsection{{\rm \footnotesize K-BAND METALLICITY DEPENDENCE}}
\label{s-met}
An argument may be made that the trends displayed in Figs.~\ref{fig-pos} and \ref{fig-r} are tied to a (radial) metallicity gradient.  Existing estimates concerning the impact of metallicity on $K$-band distances display a spread \citep[e.g.,][and references therein]{fe11}, although empirical estimates tend to favour a small dependence.  The impact of metallicity on distances to RR Lyrae variables as inferred from $K$-band photometry is now assessed to complement existing work with an independent determination \citep[e.g.,][]{so06}.

{\it K}$_s$ photometry was acquired as part of the Araucaria project for 78 RR Lyrae variables in the Sculptor dwarf spheroidal galaxy \citep{pi08}. The photometry can be correlated with metallicities established for RR Lyrae variables in that galaxy using VLT spectra \citep{cl05}.  The metallicity for each RR Lyrae variable was determined by comparing the H and Ca II K line strengths against a calibration tied to variables in globular clusters \citep[][and references therein]{cl05}.  A total of 29 variables were found to be in common between the \citet{pi08} and \citet{cl05} datasets.  Such data are relevant since the comparatively small stellar density of the dSph galaxy reduces biases introduced by crowding/blending.  The significant metallicity spread also permits an evaluation of the impact of that variable on distance.  The following expression was inferred from the data after: the removal of two outliers, considering uncertainties in $K_s$, considering a global $0.15$ dex uncertainty in ${\rm[Fe/H]}_{ZW}$ \citep{cl05}, and adopting a fixed coefficient for the pulsation term ($-2.3$): 
\begin{eqnarray}
\nonumber
K_s=(0.059\pm0.095){\rm[Fe/H]}_{ZW}-2.3\log{P_0}+(18.51\pm0.18)
\end{eqnarray}
The results support prior findings that the {\it K}-band period-magnitude relation for RR Lyrae variables is comparatively insensitive to variations in chemical composition \citep[Fig.~\ref{fig-met}, see also][and references therein]{so06,bo09,fe11}.  Consequently, the trends displayed in Figs.~\ref{fig-pos} and \ref{fig-r} are probably not tied to metallicity.  Numerous additional RR Lyrae variables in the Sculptor dSph that possess metallicity estimates \citep{cl05} do not have $K$-band observations.  Acquiring near-infrared photometry for those stars will reduce uncertainties associated with the aforementioned determinations.  Moreover, a holistic approach to constraining the impact of metallicity on RR Lyrae and Cepheid distances is preferred \citep[][and references therein]{ma11b}, rather than relying solely on individual metallicity estimates for RR Lyrae variables in a given galaxy.  A comprehensive analysis is beyond the scope of the present study since multiple lines of evidence should be employed \citep[][and references therein]{ma11b,fe11,mat12} to assess such an important correlation.

\section{{\rm \footnotesize SUMMARY}}
{\it K}-band ($2.2 \mu m$) photometry for RR Lyrae variables in the globular clusters M5, Reticulum, M92, $\omega$ Cen, and M15 exhibit a dependence on clustercentric distance. Contamination by neighbouring stars causes RR Lyrae variables near the cluster centres to appear spuriously brighter, and hence the distance moduli established for the clusters are underestimated (Figs.~\ref{fig-pos}, \ref{fig-acs}, \ref{fig-r}). That bias in turn introduces a systematic shift in the cluster ages, and by consequence on lower-limit estimates for the age of the Universe.  The results are probably not tied to variations in chemical composition among RR Lyrae variables since the {\it K}-band period-magnitude relation is relatively insensitive to that parameter (Fig.~\ref{fig-met}), as inferred from RR Lyrae variables in the Sculptor dSph.  The impact of photometric contamination may be reduced in part by an awareness of the trends exhibited in Figs.~\ref{fig-pos}, \ref{fig-acs}, and \ref{fig-r} \citep[see also][]{ya94,st05,ma12}.  Blending/crowding may be the principal (yet not the sole) source driving the clustercentric trends \citep{ma12}.

Similar results concerning the impact of photometric contamination on $W_{VI_c}$-based RR Lyrae distances were described by \citet{ma12} (Fig.~\ref{fig-pos}).  The radial trends observed are not unique to a given dataset cited, but may be present (to an extent) in most photometry sampling the cores of globular clusters.  As echoed by \citet{ma12}, the principal impetus of the present research is to increase awareness of contaminated clustercentric photometry in order to reduce the propagation of biases into research on RR Lyrae variables, especially in the era of precision cosmology where identifying and reducing systematic uncertainties is an important pursuit. For example, the Reticulum is particularly important since it has the potential to yield an accurate distance to the LMC from its rich population of RR Lyrae variables \citep{da04,wa11}.  The distance to the LMC remains a cornerstone for efforts to establish precise cosmological parameters \citep[e.g.,][]{ri11}.  Yet establishing unbiased distances to RR Lyrae variables in Reticulum appears challenging, despite (admirably) small random uncertainties associated with the photometric data (Fig.~\ref{fig-r}). 

The trends followed by the RR Lyrae variables likely promulgate into near-infrared colour-magnitude diagrams and hinder efforts to establish proper isochrone fits \citep[see][their Fig.~3]{ma12}. The evolutionary morphology displayed for cluster stars near the core differs from that exhibited by stars near the cluster periphery \citep[see also][]{ya94,st05}.  A degeneracy emerges because the signatures of multiple populations, (putative) colour-gradients, and differential reddening may be mimicked by photometric contamination.  As demonstrated here, the acquisition of multi-epoch {\it VI}$_c${\it JHK}$_s$ photometry for cluster RR Lyrae variables, which can be used to establish colour excesses and reddening-free distance moduli that are comparatively insensitive to metallicity \citep[see Fig.~\ref{fig-met}/\S \ref{s-met}, and][and references therein]{ma11b,ma12,fe11}, offers one potential solution for breaking that degeneracy.  The advent of new multi-epoch {\it JHK}$_s$ surveys such as the UKIDSS and VVV \citep{mi10,ca11,sa12}, which are monitoring globular clusters in both hemispheres, will facilitate that effort \citep[see also][]{ma12}.  Yet even higher-resolution observations, analogous to data from the HST ACS Galactic Globular Cluster Survey, may ultimately be needed to provide a comprehensive characterization of the near-infrared data discussed here.

\subsection*{{\rm \scriptsize ACKNOWLEDGEMENTS}}
\scriptsize{DM is grateful to the following individuals and consortia whose efforts and advice enabled the research: C. Clement, N. Samus, M. Catelan, G. Clementini, G. Pietrzy{\'n}ski, A. Walker, ACS Galactic Globular Cluster Survey (A. Sarajedini), CDS, arXiv, and NASA ADS.  WG is grateful for support from the BASAL Centro de Astrofisica y Tecnologias Afines (CATA) PFB-06/2007.}

\end{document}